\documentclass[sigconf]{acmart}
\usepackage{balance}
\usepackage{subcaption}
\usepackage{multirow}
\usepackage{bm}

\copyrightyear{2023} 
\acmYear{2023} 
\setcopyright{acmcopyright}\acmConference[WSDM '23]{Proceedings of the Sixteenth ACM International Conference on Web Search and Data Mining}{February 27-March 3, 2023}{Singapore, Singapore}
\acmBooktitle{Proceedings of the Sixteenth ACM International Conference on Web Search and Data Mining (WSDM '23), February 27-March 3, 2023, Singapore, Singapore}
\acmPrice{15.00}
\acmDOI{10.1145/3539597.3570384}
\acmISBN{978-1-4503-9407-9/23/02}

\begin{document}

\title{Directed Acyclic Graph Factorization Machines for CTR Prediction via Knowledge Distillation}


\author{Zhen Tian}
\affiliation{%
  \institution{Gaoling School of Artificial Intelligence, Renmin University of China}
  \city{Beijing}
  \country{China}
}
\email{chenyuwuxinn@gamil.com}

\author{Ting Bai$^{*\ddagger}$}
\affiliation{%
  \institution{Beijing University of Posts and Telecommunications}
  \city{Beijing}
  \country{China}
}
\email{baiting@bupt.edu.cn}

\author{Zibin Zhang\\ Zhiyuan Xu}
\affiliation{%
  \institution{Weixin Open Platform, Tencent}
  \city{Guangzhou}
  \country{China}
}
\email{bingoozhang@tecent.com}
\email{zhiyuanxu@tecent.com}

\author{Kangyi Lin}
\affiliation{%
  \institution{Weixin Open Platform, Tencent}
  \city{Guangzhou}
  \country{China}
}
\email{plancklin@tecent.com}


\author{Ji-Rong Wen$^\dagger$}
\affiliation{%
  \institution{Gaoling School of Artificial Intelligence, Renmin University of China}
  \city{Beijing}
  \country{China}
}
\email{jrwen@ruc.edu.cn}

\author{Wayne Xin Zhao$^\dagger$}
\affiliation{%
  \institution{Gaoling School of Artificial Intelligence, Renmin University of China}
  \city{Beijing}
  \country{China}
}
\email{batmanfly@gmail.com}

\thanks{$^*$ Ting Bai (baiting@bupt.edu.cn) is the corresponding author.}
\thanks{$^\ddagger$ Also with Beijing Key Laboratory of Intelligent Telecommunications Software and Multimedia}
\thanks{$^\dagger$ Also with Beijing Key Laboratory of Big Data Management and
Analysis Methods}




%

\newcommand{\ie}{\emph{i.e.,} }
\newcommand{\eg}{\emph{e.g.,} }
\newcommand{\paratitle}[1]{\vspace{1.5ex}\noindent\textbf{#1}}

\newcommand{\tian}[1]{\textcolor{blue}{#1}}
\newcommand{\modified}[1]{\textcolor{pink}{#1}}

\renewcommand{\authors}{Zhen Tian, Ting Bai, Zibin Zhang, Zhiyuan Xu, Kangyi Lin, Ji-Rong Wen and Wayne Xin Zhao}
\renewcommand{\shortauthors}{Zhen Tian et al.}

\begin{abstract}
With the growth of high-dimensional sparse data in web-scale recommender systems, 
the computational cost to learn high-order feature interaction in CTR prediction task largely increases, which limits the use of high-order interaction models in real industrial applications. Some recent knowledge distillation based methods transfer knowledge from complex teacher models to shallow student models for accelerating  the online model inference. However, they suffer from the degradation of model accuracy in knowledge distillation process. It is challenging to balance the efficiency and effectiveness of the shallow student models.
To address this problem, we propose a \textbf{D}irected \textbf{A}cyclic \textbf{G}raph \textbf{F}actorization \textbf{M}achine (\textbf{KD-DAGFM}) to learn the high-order feature interactions from existing complex interaction models for CTR prediction via \textbf{K}nowledge \textbf{D}istillation.
The proposed lightweight student model \textbf{DAGFM} can learn arbitrary explicit feature interactions from teacher networks, which achieves approximately lossless performance and is proved by a dynamic programming algorithm.
Besides, an improved general model KD-DAGFM+ is shown to be effective in distilling both explicit and implicit feature interactions from any complex teacher model.
Extensive experiments are conducted on four real-world datasets, including a large-scale industrial dataset from WeChat platform with billions of feature dimensions. KD-DAGFM achieves the best performance
with less than $21.5\%$ FLOPs of the state-of-the-art
method on both online and offline experiments, showing the superiority of DAGFM to deal with the industrial scale data in CTR prediction task\footnote{Our implementation code is available at: \url{https://github.com/RUCAIBox/DAGFM}}.

\end{abstract}



\maketitle

\section{Introduction}
Click-Through Rate (CTR) prediction, which aims to predict the probability of a user clicking on an item, is a very critical task in recommender systems. 
The key to achieve good performance in CTR prediction is learning the effective high-order feature interactions, which has attracted great attention in recent years~\cite{cheng2016wide,he2017neural,li2020interpretable,shan2016deep}. 
One of the biggest challenges is the high computational cost to model the high-order interactions of raw features, because the number of feature combinations increases exponentially when the number of raw features increases. 
In real-world applications, raw features are usually highly sparse with millions of dimensions. 
For example, identifier feature such as the ID of user/item is very sparse after being encoded as an one-hot vector; so are the multi-filed vectors built from upstream tasks such as visual information.
It is very time-consuming to calculate the high-order feature interactions on such sparse features with millions of dimensions, and there is a high risk of over-fitting problem for CTR prediction in real industry recommender systems.

Hence, a lightweight recommendation algorithm for CTR prediction is needed to simplify the online inference process, which enables it to avoid the explosion of computational costs in real industry recommender systems. 
Some efforts have been made to solve this problem~\cite{zhou2018rocket,zhu2020ensembled,xu2020privileged}. They utilize Knowledge Distillation (KD) technique to transfer knowledge from complex teacher models to shallow student models with reduced learning parameters, so as to speed up the model inference for dealing with the real-time massive data in recommender systems.
However, the acceleration of KD model with reduced learning parameters is at the expense of the degradation of the model accuracy, and it is inevitable for KD model to balance the effectiveness and efficiency~\cite{zhou2018rocket}.
The KD method~\cite{zhu2020ensembled} achieves better performance than its teacher model, but at cost of an undiminished amount of parameters of the student model, which may limit its adoption in online inference process. 

To keep the low model complexity and meanwhile achieve better performance of KD student model, we propose a lightweight \textbf{K}nowledge \textbf{D}istillation based \textbf{D}irected \textbf{A}cyclic \textbf{G}raph \textbf{F}actorization \textbf{M}achine (\textbf{KD-DAGFM}) to predict the CTR in recommender systems. 
Our aim is to design a shadow student model with fewer learning parameters, but maintain the capability to distill the high-order feature interactions from existing complex interaction models. 
To achieve the approximately lossless knowledge distillation, we carefully design the student model: Directed Acyclic Graph Factorization Machine (\textbf{DAGFM}) to learn arbitrary explicit feature interactions. 
Specifically, each node in Directed Acyclic Graph (DAG) represents a feature field, and different nodes are connected by directed edges without cycles. 
The interaction of nodes, which represents the interaction of features, are modeled by the learnable weights of edges in DAG.
With the specific interaction learning functions (\ie inner, outer and kernel), the feature interactions on DAG can be translated into explicit arbitrary-order interactions, which is demonstrated by a dynamic programming algorithm. 
Except for the explicit feature interactions, to distill the knowledge from implicit feature interaction models (\eg AutoInt~\cite{song2019autoint} and FiBiNet~\cite{huang2019fibinet}), we further propose an improved model KD-DAGFM+, in which an Multi-Layer Perceptron (MLP) component is integrated into the student model DAGFM. 
Experiments show that KD-DAGFM+ has the capability to distill both explicit and implicit feature interactions from any complex teacher models. 
The contributions are summarized as follows:



\begin{itemize}
\item 
We propose a lightweight distillation model KD-DAGFM to learn high-order explicit interactions from different complex teacher models, which could greatly reduce the computational costs (\ie FLOPs) by at least 10 times, making it possible to be applied into large-scale industry recommender systems.

\item 
We design a directed acyclic graph neural network -- DAGFM as the student model and use three different interaction learning functions (\ie inner, outer and kernel functions) in the propagation process to capture arbitrary-order explicit feature interactions, which achieves approximately lossless performance of the student model and is demonstrated by a dynamic programming algorithm.

\item KD-DAGFM achieves the best performance 
with less than $21.5\%$ FLOPs of the state-of-the-art method on both online and offline experiments, showing the superiority of KD-DAGFM to deal with the large scale industry data in CTR prediction task.
\end{itemize}

\section{Preliminary} \label{sec:exh}
We first give a preliminary introduction to the CTR prediction task, then introduce the explicit high-order feature interaction and the interaction learning function in CTR prediction task. 

\subsection{CTR Prediction}
Click-Through Rate (CTR)  prediction aims to predict the probability of the user clicking on an item. Specifically, given the $m$ feature fields, we use 
$\bm x = \{\bm x_1, \bm x_2,...,\bm x_m \}$ to represent the feature of users and items, where $\bm x_i$ is the feature representation of the $i$-th feature. Label $y \in \{0 , 1\}$  represents each item is click or not, and is predicted based on the input feature $\bm x$.
The key to achieve good performance in CTR prediction task is learning the effective high-order feature interactions, which is a fundamental problem and has attracted great attention in research areas.

\subsection{High-order Feature Interactions}
Given the input feature $\bm x = \{\bm x_1, \bm x_2,...,\bm x_m \}$, the embedding features of $\bm x$ is represented as
$\{ \bm{e}_1,\bm{e}_2,..,\bm{e}_m \}$, where $m$ is the number of feature fields, the explicit feature interactions can be formulated as:
\begin{equation} \label{eq:fi}
  \hat{y}=
  \sum_{t = 2}^m \sum_{j_1 < j_2 < ... < j_t}
  \phi (\bm e_{j_1} , \bm e_{j_2} , ... ,\bm e_{j_t} ),
\end{equation}
where $\phi$ denotes the feature interaction learning function.

Take the $2$-order feature interactions as example, formulated as:
\begin{equation}
  \hat{y}=
  \sum_{i = 1}^m \sum_{j = i+1}^m
  \phi (\bm e_{i} , \bm e_{j}).
\end{equation}


\subsection{Interaction Learning Function}
The feature interaction learning function defines the way to compute the interactions among features. 
Different interaction models utilize different interaction learning functions. 



\begin{figure*}[htbp]
\centering
\begin{subfigure}{.49\textwidth}
  \centering
  \includegraphics[width=.6\textwidth]{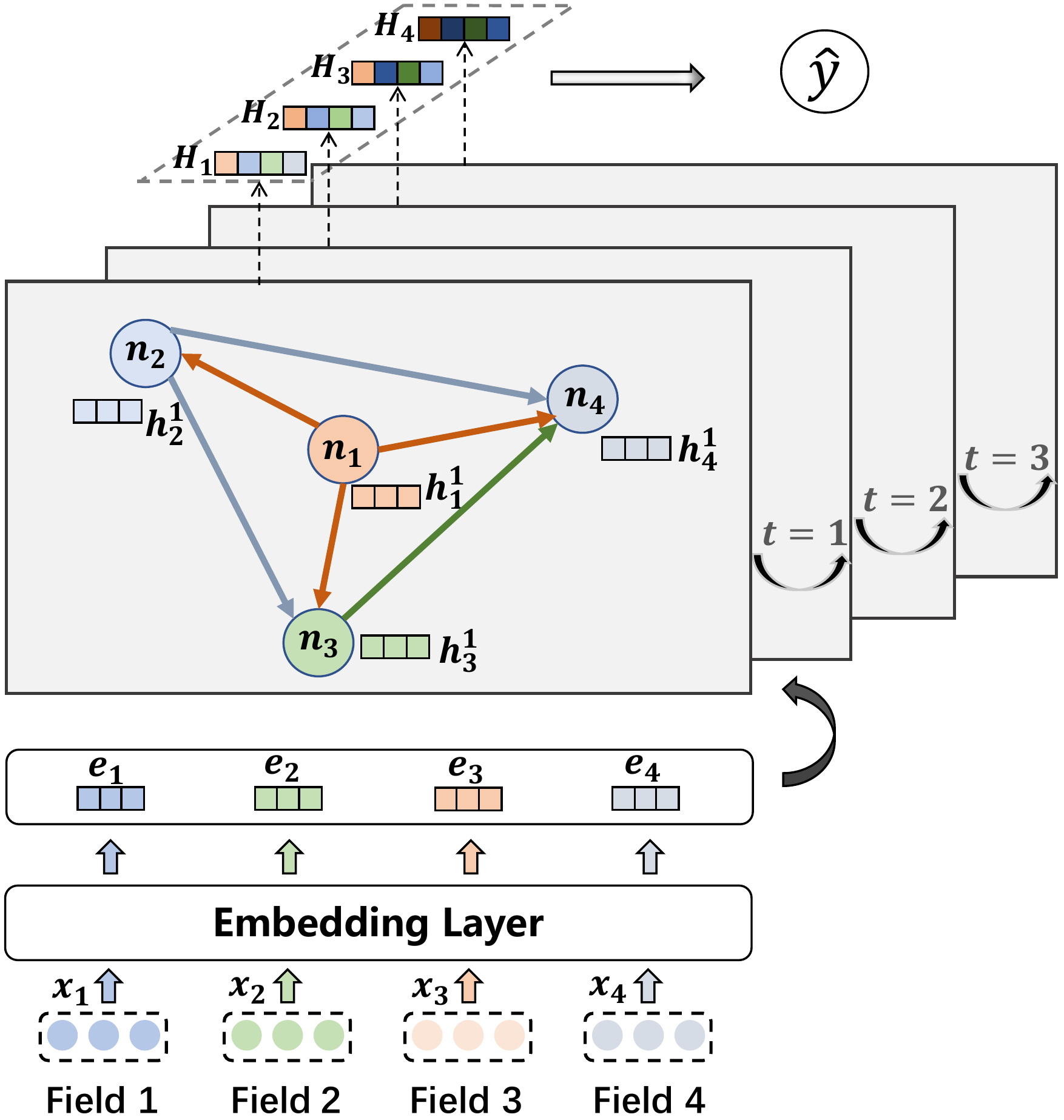}
  \caption{The architecture of DAGFM.}
\end{subfigure} \hfill  
\begin{subfigure}{.5\textwidth}
  \centering
  \includegraphics[width=1.\textwidth]{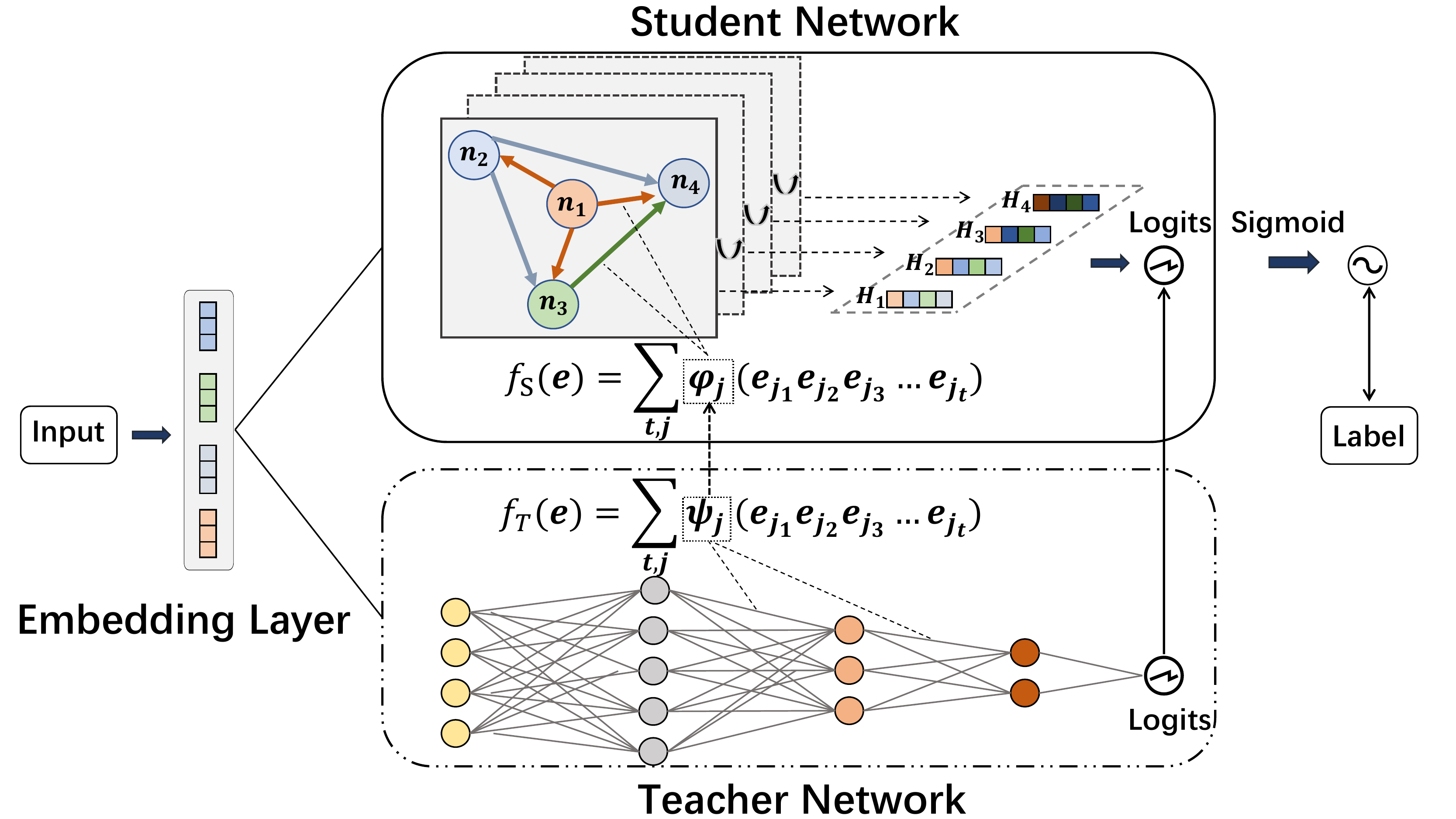}
  \caption{The architecture of KD-DAGFM.}
\end{subfigure} 
\captionsetup{font={small}}
\caption{Components and architecture of our proposed KD-DAGFM.}
\label{fig:framework}
\end{figure*}

\subsubsection{Basic Inner Interaction Learning Function}
The most commonly used interaction learning function in existing studies, \ie FM~\cite{rendle2010factorization}, IM~\cite{yu2020deep}, HOFM~\cite{blondel2016higher}, is the basic inner interaction learning function, and it can be formulated as:
\begin{equation}
\label{eq:simple}
\phi (\bm e_i, \bm e_j)= \bm e_i \odot \bm e_j ,
\end{equation}
where $\odot$ is the element-wise product of two vectors.

\subsubsection{Weighted Inner Interaction Learning Function}
The above methods with basic inner function can not capture the distinct field information of features, leading to gradient coupled issue~\cite{qu2018product}. 
To address this problem, many field-aware variants such as FwFM~\cite{pan2018field} and FvFM~\cite{sun2021fm2} use weight parameters to model the feature interactions of different fields. The inner interaction learning function used in them can be formulated as:
\begin{equation}
\phi (\bm e_i, \bm e_j)= \bm w_{i,j} \odot \bm e_i \odot \bm e_j,
\label{eq:in}
\end{equation}
where $\bm w_{i,j}$ is a weight vector between the fields $i$ and $j$. 

\subsubsection{Kernel Interaction Learning Function}
By extending the dimension of weight vector, recent studies~\cite{qu2018product,sun2021fm2} propose a more powerful interaction learning function, \ie kernel interaction learning function, defined as:
\begin{equation}
  \phi(\bm e_i , \bm e_j) = \bm e_i \bm W_{i , j}  \odot \bm e_j,
  \label{eq:kernel}
\end{equation}
where $\bm W_{i , j}$ is the learning weight matrix.

Despite the great improvements by incorporating field weights, the model complexity has largely increased.
Enumerating all high-order feature interactions with distinct field weights is an NP-hard problem. 
Most existing high-order feature interaction models~\cite{lian2018xdeepfm,wang2021dcn} improve the model capability by increasing the parameters, but at the expense of high computational costs, making it impractical to be applied into large-scale industry recommendation scenarios.


\section{Methodology}


To improve model performance with low computational costs, we propose a Knowledge Distillation based Directed Acyclic Graph Factorization Machine (KD-DAGFM),
in which a lightweight student model Directed Acyclic Graph Factorization Machine (DAGFM) is designed to learn the high-order feature interactions from existing complex teacher models.

 
\subsection{Directed Acyclic Graph Factorization Machine (DAGFM)} \label{sec:LightGFM}

To effectively and efficiently extract knowledge from complex high-order interaction models, we design a lightweight student model, named \textbf{D}irected \textbf{A}cyclic \textbf{G}raph \textbf{F}actorization \textbf{M}achine (\textbf{DAGFM}).
The architecture of DAGFM is shown in Fig.~\ref{fig:framework} (a), we first introduce the construction of Directed Acyclic Graph (DAG) and the information interaction process, then 
introduce
our proposed efficient outer interaction learning function.


\subsubsection{The Construction of DAG}
As shown in Fig.~\ref{fig:framework}(a), 
given the input features $\bm x = \{\bm x_1,\bm x_2,...,\bm x_m\}$, where $m$ is the number of total feature fields, and $\bm x_i$ is the feature representation of the $i$-th feature.
The embedding layer projects the input features from a high-dimensional sparse space to a lower-dimensional dense space, \ie $\bm e_i = \bm x_i \bm W_i $, where $\bm W_i \in \mathbb{R}^{|\bm{x}_i| \times d}$ and $d$ is the embedding size.

The directed acyclic graph (DAG) is defined as $\mathcal{G}=(\mathcal{N} , \mathcal{E})$, where each node $n_i \in \mathcal{N}$ corresponds to a feature field and $|\mathcal{N}|=m$. The initial state vector of each node $n_i$ is set to $\bm e_i$, \ie $\bm h_{i}^1=\bm e_i$.
The edge from node $n_i$ to $n_j$ ($j>i$) is directed, which controls the interaction weight between two features from different fields.

\subsubsection{Learning Feature Interactions on DAG}
The interactions among features can be modeled as the information propagation process on DAG in a recurrent fashion. 
At each propagation layer, each node will aggregate the state from its neighbors and combine it with the initial state $\bm h_i^1$ of itself to update the state vector.
Formally, the state vector of node $n_i$ at propagation layer $t$ can be represented as:
\begin{equation} \label{eq:agg}
  \bm h_i^{t+1} = \sum_{j \in \mathcal{N}(i)\cup \{i\}} { \phi (\bm h_j^t,\bm h_i^1) },
\end{equation}
where $\mathcal{N}(i)=\{j | n_j \rightarrow n_i \in \mathcal{E}\}$, and $\phi$ is the interaction learning function.

To obtain the information of high-order feature interactions for CTR prediction, we first apply sum pooling of the hidden state vector for each node as ${p_i^t}  = \sum_{k = 1}^{d}{h_{i,k}^t}$ at each interaction step, where $h_{i,k}^t$ denotes the $k$-th element of $\bm h_i^t$.
Then we concatenate all node statesp at layer $t$ as $\bm p^t = [p_1^t,p_2^t,...,p_m^t]$, and the different-order interactions of features on DAG can be represented as $\bm p = [\bm p^1;\bm p^2;...;\bm p^{l}]$, where the number of propagation layers is $l-1$. 
Finally, the prediction function for CTR prediction can be formulated as: 
\begin{equation}
  \hat{y} = \sigma(\bm p \bm w^\top + b),
\end{equation}
where $\sigma$ is the sigmoid activation function, $\bm{w}$ is the transition vector and $b$ is the bias.

\subsubsection{Outer Interaction Learning Function}
The feature interaction learning function $\phi$ of DAGFM (see in Eq.~\ref{eq:agg}) is flexible, we can utilize the most commonly used learning functions, \ie the inner interaction learning function (see Eq.~\ref{eq:in}) or the kernel interaction learning function with more powerful model capability (see Eq.~\ref{eq:kernel}), to learn the high-order feature interactions. 

To reduce the computational complexity of kernel interaction learning function, we propose a simplified \textbf{outer interaction learning function},
in which the learning weight matrix in kernel interaction learning function (see Eq.~\ref{eq:kernel}) can be decomposed as: 
\begin{equation}
    \bm W_{j , i}^t = (\bm p_{j , i}^t)^\top  \bm q_{j , i}^t,
\end{equation}
where $\bm p_{j , i}^t \in \mathbb{R}^{d}$ and $\bm q_{j , i}^t \in \mathbb{R}^{d}$ are the decomposed vectors.

Then the outer interaction learning function can be defined as:
\begin{equation}
\phi(\bm h_j^t,\bm h_i^1)  = \bm h_j^t \bm W_{j,i}^t \odot \bm h_i^1 = (\bm h_j^t  (\bm p_{j , i}^t)^\top) \cdot (\bm q_{j , i}^t \odot \bm h_i^1).
  \label{eq:outer}
\end{equation}

Our proposed outer interaction learning function is decomposed from kernel interaction learning function (see Eq.~\ref{eq:kernel}) and has the same complexity with inner interaction learning function (\ie $O(d)$), which is much less than kernel interaction learning function (\ie $O(d^2)$), making it more efficient to capture the feature interactions.

\subsection{Arbitrary-Order Interaction Learning in DAGFM}
\label{sec:fi ana}
In this section, we demonstrate that the feature interactions learned in DAGFM align with different unique weighted paths in a dynamic programming (DP) algorithm, showing the ability of DAGFM to learn arbitrary-order feature interactions.

We define the suffix of a feature interaction as the feature with the largest subscript.
For example, for the interactions with feature $\bm e_2 \bm e_3$ and $\bm e_3 \bm e_4 \bm e_5$, the suffix are $\bm e_3$ and $\bm e_5$ respectively.
Let $\bm S_i^{t}$ denote the set of all $t$-th order feature interactions suffixed with feature $\bm e_i$. 
Take 2-order feature interactions  suffixed with $\bm{e}_3$ as example: $\bm S_3^{2} = \{\bm e_{1} \bm e_{3} , \bm e_{2} \bm e_{3},\bm e_{3} \bm e_{3}\}$. The 1-order feature interaction suffixed with $\bm{e}_i$ is itself, \ie $\bm S_i^{1} = \{\bm e_{i}\}$.

In dynamic programming algorithm, the state variable
$\bm h_i^{t}$ is defined as the sum of elements in $\bm S_i^{t}$, \ie $\bm h_i^{t} = \sum_{\bm{j} \in \bm S_i^{t}}{\bm{j}}$.  
In fact, all $t+1$-th order interactions suffixed with $\bm e_i$ can be expressed as the interaction between $\bm e_i$ and $t$-th order interactions suffixed with $\bm e_j$ ($j \leq   i$).
Therefore, the state transfer equation of $\bm h_i^{t}$ in DP algorithm can be formulated as:
\begin{equation} \label{eq:trans}
  \bm h_i^{t+1} = \sum_{j=1}^{i} {\bm h_j^{t} \odot \bm e_i} = \sum_{j=1}^{i} {\bm h_j^{t} \odot \bm h_i^1}.
\end{equation}


In DAGFM, the feature interactions are propagated based on a directed acyclic graph, which means the neighbors $\mathcal{N}(i)$ of node $n_i$ refer to the directly linked nodes $n_j$, where $j < i$. 
Hence, the state vector of node $n_i$ in graph propagation process in Eq.~\ref{eq:agg} is equivalent to the transferred state in DP algorithm (see Eq.~\ref{eq:trans}).  
\begin{equation} \label{eq:eq}
  \bm h_i^{t+1} = \sum_{j \in \mathcal{N}(i)\cup \{i\}} { \phi (\bm h_j^t,\bm h_i^1) }
  = \sum_{j=1}^{i} \bm h_j^t \odot \bm h_i^1,
\end{equation}
here the interaction learning function $\phi$ is equal to the basic inner interaction learning function (see in Eq.~\ref{eq:simple}). 

\begin{figure}[!t]
  \centering
  \includegraphics[width=0.9\linewidth]{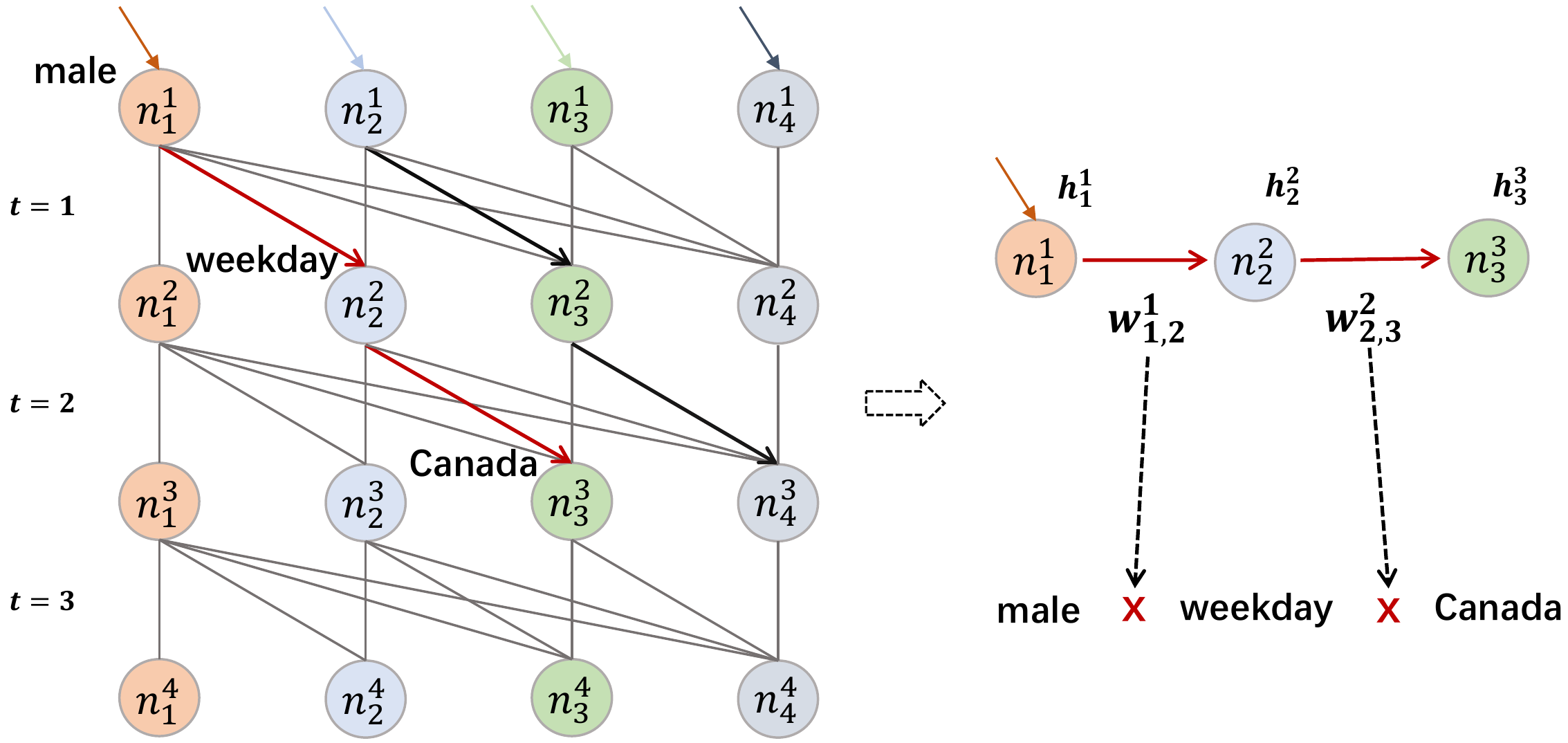}
  \captionsetup{font={small}}
  \caption{The propagation graph of DAGFM. Each $k$-order feature interaction corresponds to a unique path with length $k-1$ in the dynamic programming algorithm.}
  \label{fig:Prop}
\end{figure}

From Eq.~\ref{eq:trans} and Eq.~\ref{eq:eq}, we can see that each feature interaction in DAGFM can align with a unique path in DP graph (see in Fig.~\ref{fig:Prop}): each $k$-order feature interaction corresponds to a unique path with length $k-1$ from the first layer. 
By summing all the state vectors in DP algorithm, all the paths can be traversed, indicating all the feature interactions can be modeled in DAGFM.

For graph propagation processes with inner (see Eq.~\ref{eq:in}) and kernel (see Eq.~\ref{eq:kernel}) interaction learning functions, they can be equivalent to Eq.~\ref{eq:eq} by assigning the weights $\bm w_{j , i}^t$ to be all-ones vector and identity matrix respectively. Besides, the weights can also be learned during training process, which enables DAGFM to model all types of feature interactions with different semantic information. 
For example, let nodes $n_1,n_2,n_3,n_4$ represent \verb|male|, \verb|weekday|, \verb|Canada|, \verb|sunny| at first layer respectively.
Feature interaction \verb|male| $\times$ \verb|weekday| $\times$ \verb|Canada| corresponds to the path $n_1^1 \rightarrow n_2^2 
\rightarrow n_3^3$ marked red color with the learning weights $\bm w_{1,2}^1 \odot \bm w_{2,3}^2$. 
The edge weight $\bm w_{i,j}^t$ controls the strength of the interaction between $i$-th feature and 
$j$-th feature at $t$-th order interactions. 
In the above example, $\bm w_{1,2}^1$ controls the semantic strength of \verb|male| $\times$ \verb|weekday| in \verb|male| $\times$ 
\verb|weekday| $\times$ \verb|Canada|.

\subsection{Knowledge Distillation Process} \label{sec:approach}

The overall distillation framework of KD-DAGFM is shown in Fig.~\ref{fig:framework}(b). 
In knowledge distillation process, the student network DAGFM learns the knowledge of feature interactions from teacher networks.
To ensure the feature space in student and teacher networks be the same, 
we share the embedding layer of  teacher network with student network.
The loss function in knowledge distillation can be formulated as:
\begin{equation}
    \mathcal{L}_{KD} = \frac{1}{N} \sum_{i=1}^{N}\bigg(T(\bm x_i,\bm{\psi}) - S(\bm x_i,\bm{\varphi})\bigg)^2,
  \end{equation}
where $N$ is the total number of training instances, $T$ and $S$ denote the learning function, $\bm{\psi}$ and $\bm{\varphi}$ are the parameters in teacher and student networks respectively.

Besides, we adopt cross entropy loss function for CTR prediction, which could be formulated as:
  \begin{equation}
    \mathcal{L}_{CTR} = -\frac{1}{N} \sum_{i=1}^{N}\bigg(y_i\log(\hat{y}_i)+(1-y_i)\log(1-\hat{y}_i)\bigg),
  \end{equation}
where $y_i$ is the true label ,$\hat{y}_i$ is predicted result.

The final optimization objective of the knowledge distillation model KD-DAGFM is:
  \begin{equation}
    \mathcal{L}= \alpha \mathcal{L}_{KD} + \beta \mathcal{L}_{CTR},
  \end{equation}
where $\alpha , \beta$ are hyper-parameters to balance the weights of two loss functions.

After knowledge distillation, the student model can well learn the useful feature interactions from the teacher network. 
However, since the feature embedding of student is inherited from teacher which is not trained during distillation, it may result in the under-fitting problem and make sub-optimal performance of the student model. To alleviate this problem, we fine-tune the student model for the final CTR prediction. 

\section{EXPERIMENTS}
We conduct experiments to demonstrate the effect of our proposed knowledge distillation model KD-DAGFM.
Four real world datasets are used in our experiments: Criteo\footnote{http://labs.criteo.com/2014/02/kaggle-display-advertising-challenge-dataset}, Avazu\footnote{http://www.kaggle.com/c/avazu-ctr-prediction}, MovieLens-1M\footnote{https://grouplens.org/datasets/movielens} and WeChat. 
Criteo is the most popular CTR prediction benchmark dataset contains user logs over a period of 7 days. 
Avazu contains user logs over a period of 7 days, which was used in the Avazu CTR prediction competition.
MovieLens-1M is the most popular dataset for recommendation systems research. 
WeChat is collected on WeChat platform, which contains more than one hundred million daily active users.
Table~\ref{tab:datasets} summarizes the dataset statistics information. 

We first verify the effectiveness of our proposed student model DAGFM and then compare the model performance of KD-DAGFM with the state-of-the-art interaction models for CTR prediction task.

\begin{table}[]
  \captionsetup{font={small}}
  \caption{The statistics of datasets.} 
  \label{tab:datasets}
  \resizebox{.32\textwidth}{!}{
  \begin{tabular}{c|ccc}
    \toprule
    \textbf{Dataset}& $\#$ Features & $\#$ Fields & $\#$ Instances \\
    \hline \hline
    Criteo & 1.3M & 39 & 45M \\
    Avazu & 1.5M & 23 & 40M \\
    MovieLens-1M & 13k & 7 & 740K\\
    WeChat & 2.9M & 264 & 40.9M \\
  \bottomrule
\end{tabular}}

\end{table}


\subsection{Effectiveness Analysis of the Student Model DAGFM}
In this section, we conduct experiments to demonstrate that our proposed student model DAGFM  can achieve the approximately lossless performance of the complex teacher models.
    

\subsubsection{The Teacher Networks}  We choose the most commonly used complex models with competitive performance in CTR prediction task, \ie  xDeepFM~\cite{lian2018xdeepfm} and DCNV2~\cite{wang2021dcn}. 
Due to the fact that DAGFM is an explicit model and its distillation from deep implicit component may incorporate noise to degrade model performance,
for fair comparison and speed up the knowledge distillation process, 
we only use the explicit feature interaction component \textbf{CIN} and \textbf{CrossNet} as the teacher models in xDeepFM and DCNV2 respectively.
(results with different types of teachers, including both implicit and explicit teacher networks, are shown in Sec.~\ref{sec: generalkd}.)
\subsubsection{The Student Networks}  Considering the high computational costs of deep methods may limit their adoption in large-scale recommendation scenarios, we use the most commonly used light models: \textbf{FwFM}~\cite{pan2018field}, \textbf{FmFM}~\cite{sun2021fm2}, and \textbf{tiny MLP} (\ie the MLP with only three layers: $128 \times 128 \times 128$) as the student models for comparison. For DAGFM, we adopt the best combination of teachers and interaction learning functions, \ie CIN with DAGFM-inner and CrossNet with DAGFM-outer (more detailed experimental results are analyzed in Sec.~\ref{sec: expif}).

\subsubsection{Implementation Details} The experimental settings of the knowledge distillation process is identical to the settings of KD-DAGFM in Section~\ref{exp: main}. The details of implementation are summarized in Section~\ref{sec: impd}.

\begin{table}[]
  \captionsetup{font={small}}
  \caption{Effectiveness comparisons of different student models. $l$ is the depth, $m$ is the number of feature fields, $d$ is the embedding size, $H$ is the dimension of hidden vectors. 
  }
  \label{tab:kd performance}
  \resizebox{.5\textwidth}{!}{
  \begin{tabular}{l|cc|cc|cc}
    \toprule
    \multirow{2}{*}{\textbf{ Distillation}}&\multicolumn{2}{c|}{\textbf{Criteo}}&\multicolumn{2}{c|}{\textbf{Avazu}} & \multirow{2}{*}{\textbf{Order}} & \multirow{2}{*}{\textbf{Complexity}}\\
    & AUC & Log Loss  & AUC & Log Loss \\
    \hline \hline
    CIN & 0.8109 & 0.4424 &  0.7816 & 0.3803 & $\geq 2$ & $O(mH^2dl)$\\
    CIN $\rightarrow$  FwFM  & 0.8008 & 0.4511  & 0.7779 & 0.3823 & $2$ & $O(m^2d)$ \\
    CIN $\rightarrow$ FmFM  & 0.8091 & 0.4445  & 0.7809 & 0.3806 & $2$ & $O(m^2d^2)$ \\
    CIN $\rightarrow$ Tiny MLP  & 0.8098 & 0.4506  & 0.7794 & 0.3839 & $NA$ & $O(mdH + H^2l)$ \\
    CIN $\rightarrow$ DAGFM-inner  & \textbf{0.8109} & \textbf{0.4425} & \textbf{0.7816} & \textbf{0.3803} & $\geq 2$ & $O(m^2dl)$\\
    \hline
    CrossNet & 0.8123 & 0.4398 &  0.7817 & 0.3805 & $\geq 2$ & $O(m^2d^2l)$\\
    CrossNet $\rightarrow$ FwFM  & 0.7945 & 0.4559  & 0.7690 & 0.3874 & $2$ & $O(m^2d)$ \\
    CrossNet $\rightarrow$ FmFM  & 0.8108 & 0.4411  & 0.7800 & 0.3811 & $2$ & $O(m^2d^2)$ \\
    CrossNet $\rightarrow$ Tiny MLP  & 0.8102 & 0.4516  & 0.7795 & 0.3837 & $-$ & $O(mdH + H^2l)$ \\
    CrossNet $\rightarrow$ DAGFM-outer  & \textbf{0.8122} & \textbf{0.4397}  & \textbf{0.7815} & \textbf{0.3806} & $\geq 2$ & $O(m^2dl)$ \\
  \bottomrule
\end{tabular}}
\end{table}

\begin{figure}[htbp]
\centering
\begin{subfigure}{.5\textwidth}
  \centering
  \includegraphics[width=.85\textwidth]{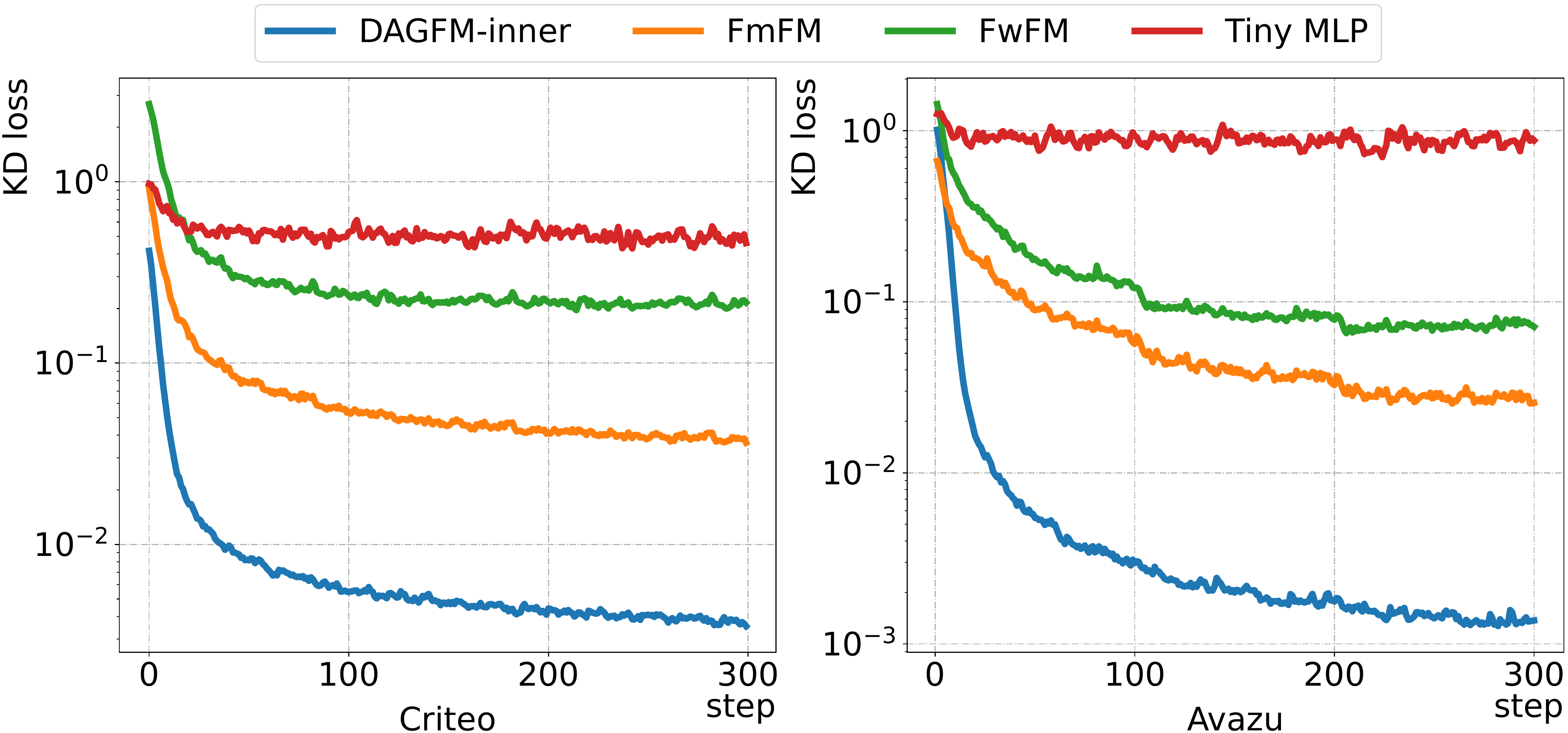}
  \caption{CIN}
\end{subfigure}
\begin{subfigure}{.5\textwidth}
  \centering
  \includegraphics[width=.85\textwidth]{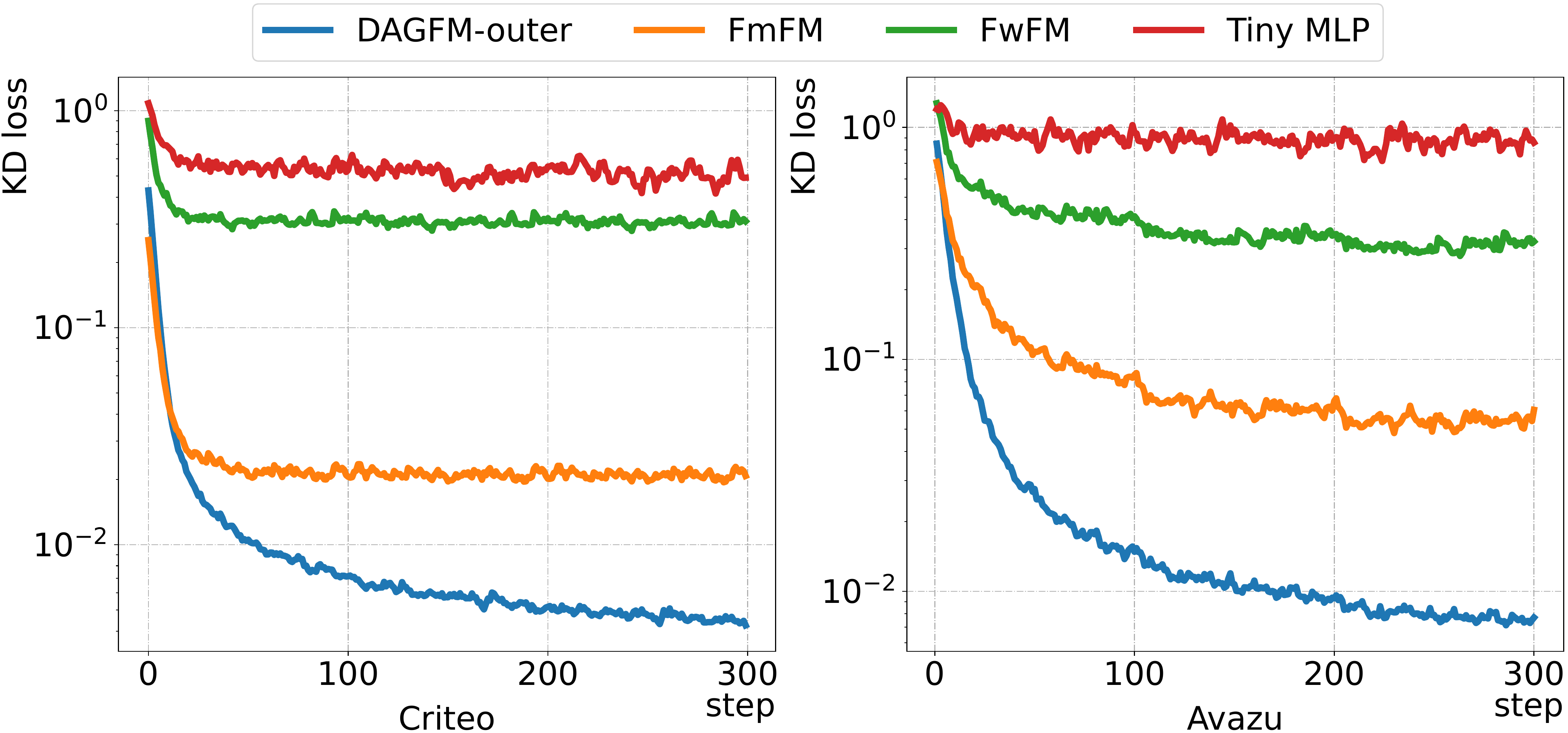}
  \caption{CrossNet}
\end{subfigure} 
\captionsetup{font={small}}
\caption{The loss curves in knowledge distillation process of different student models.}
\label{fig:loss curve}
\end{figure}

\begin{table*}[]
  \centering
  \small
  \captionsetup{font={small}}
  \caption{Performance comparisons. Note that a higher AUC or lower Logloss at 0.001-level is significant for CTR prediction.}\label{tab:overall}
  \resizebox{.9\textwidth}{!}{
  \begin{tabular}{c|cc|cc|cc|ccrrr}
    \toprule
    \multirow{2}{*}{\textbf{Model}}&
    \multicolumn{2}{c|}{\textbf{Criteo}}&\multicolumn{2}{c|}{\textbf{Avazu}}&\multicolumn{2}{c|}{\textbf{MovieLens-1M}}&\multicolumn{5}{c}{\textbf{WeChat}}\\
    &AUC&Log Loss& AUC & Log Loss & AUC & Log Loss & AUC & Log Loss & Params & FLOPs & Latency \\
    \hline\hline
    FmFM & 0.8112 & 0.4408 & 0.7744 & 0.3831 & 0.8864 & 0.3295 & 0.6593 & 0.2660 & 5.99M & 9.44M & 0.099 ms\\
    FwFM & 0.8104 & 0.4414 & 0.7741 & 0.3835 & 0.8815 & 0.3351 & 0.6702 & 0.2637 & 0.03M & 1.11M & 0.046 ms\\
    xDeepFM & 0.8122 & 0.4407 & 0.7821 & 0.3799 & 0.8913 & 0.3244 & 0.6712 & 0.2627 & 282.77M & 3761.16M & 0.588 ms\\
    DCNV2 & \underline{0.8127} & \underline{0.4394} & \underline{0.7838} & \underline{0.3782} & 0.8946 & 0.3229 & 0.6683 & 0.2640 & 87.63M & 87.63M & 0.198 ms \\
    FiBiNet & 0.8126 & 0.4415 & 0.7837 & 0.3783 & 0.8860 & 0.3291 & 0.6681 & \textbf{0.2449} & 569.01M & 587.76M & 0.219 ms\\
    AutoInt+ & 0.8126 & 0.4396 & 0.7832 & 0.3786 & 0.8937 & 0.3288 & \underline{0.6774} & 0.2618 & 34.14M & 64.92M & 0.222 ms\\
    FiGNN & 0.8109 & 0.4412 & 0.7830 & 0.3799 & 0.8939 & 0.3232 & 0.6623 & 0.2641 & 9.91M & 41.13M & 0.323 ms\\
    GraphFM & 0.8070 & 0.4448 & 0.7792 & 0.3807 & 0.8890 & 0.3311 & 0.6532 & 0.2660 & 3.60M & 1193.74M & 0.192 ms \\ 
    ECKD & 0.8123 & 0.4422 & 0.7834 & 0.3838 & \underline{0.8951} & \textbf{0.3173} & 0.6635 & 0.2672 & 25.44M & 25.44M & 0.108 ms\\
    \hline \hline
    CIN (teacher) & 0.8109 & 0.4424 & 0.7816 & 0.3803 & 0.8850 & 0.3320 & 0.6668 & 0.2636 & 231.96M & 3710.57M & 0.213 ms\\
    DAGFM-inner (student) & 0.8105 & 0.4413 & 0.7801 & 0.3805 & 0.8839 & 0.3339 & 0.6620 & 0.2651 & 1.75M & 3.36M & 0.068 ms \\
    KD-DAGFM-inner & 0.8109 & 0.4425 & 0.7816 & 0.3803 & 0.8849 & 0.3320 & 0.6668 & 0.2636 & 1.75M & 3.36M & 0.068 ms \\
    KD-DAGFM$_{FT}$-inner & 0.8121 & 0.4400 & \textbf{0.7883} & \textbf{0.3760} & 0.8880 & 0.3304 & \textbf{0.6777} & \underline{0.2617} & \textbf{1.75M} & \textbf{3.36M} & \textbf{0.068 ms} \\
    \hline
    CrossNet (teacher) & 0.8123 & 0.4398 & 0.7817 & 0.3805 & 0.8907 & 0.3474 & 0.6681 & 0.2638 & 53.54M & 53.54M & 0.125 ms\\
    DAGFM-outer (student) & 0.8119 & 0.4401 & 0.7791 & 0.3810 & 0.8895 & 0.3361 & 0.6672 & 0.2646 & 3.42M & 5.04M & 0.086 ms\\
    KD-DAGFM-outer & 0.8122 & 0.4397 & 0.7815 & 0.3806 & 0.8904 & 0.3476 & 0.6680 & 0.2638 & 3.42M & 5.04M & 0.086 ms\\
    KD-DAGFM$_{FT}$-outer & \textbf{0.8132} & \textbf{0.4390} & 0.7864 & 0.3780 & \textbf{0.8976} & \underline{0.3189} & 0.6748 & 0.2665 & \textbf{3.42M} & \textbf{5.04M} & \textbf{0.086 ms}\\
  \bottomrule
\end{tabular}}
\end{table*}

\subsubsection{The Experimental Results of Different Student Models.}
The experimental results of different student models on Criteo and Avazu datasets are shown in Table~\ref{tab:kd performance}, we can see that:

(1) FwFM and FmFM perform the worst due to the limited  $2$-order feature interaction capability, which can not well fit the high-order information of the teacher network. 

(2) Tiny MLP learns high-order feature interactions with multi-layer neural networks in an implicit manner, and its capability is limited by the shallow model architecture.

(3) DAGFM achieves competitive performance compared with the teacher network, showing its capability to transfer the high-order feature interaction information.

(4) As the student network, the model complexity of DAGFM (\ie $m^2dl$) is much lower than the teacher model CIN (\ie $mH^2dl$) and CrossNet (\ie $m^2d^2l$) ($d$ is the size of embedding and $H$ is the dimension of hidden vectors, which are much larger than the number of fields $m$).
For the student networks, the complexity of DAGFM is much lower than FmFM and Tiny MLP and is comparable with FwFM.

We also show the distillation loss in training process. As shown in Fig.~\ref{fig:loss curve}: the deviation of DAGFM from the teacher network is very small (10$^{-3}$), which is almost 1000 times smaller than it in the tiny MLP model, 30 times and 300 times smaller than it in FmFM and  FwFM respectively, showing the approximately lossless distillation capability of DAGFM.

\subsection{Experimental Results of KD-DAGFM}\label{exp: main}
We conduct experiments to show the effectiveness of our proposed knowledge distillation model KD-DAGFM in CTR prediction task. 

\subsubsection{Compared Methods}
We compare KD-DAGFM with state-of-the-art methods in CTR prediction task, including:

$\bullet$ FmFM~\cite{sun2021fm2}: it utilizes kernel product with matrix weights to capture the field information.

$\bullet$ FwFM~\cite{pan2018field}: it utilizes inner product with scalar weights to capture the field information. 

$\bullet$ xDeepFM~\cite{lian2018xdeepfm}: it utilizes inner product to generate multiple feature maps, which encode all pairwise interactions to model explicit interactions. 

$\bullet$ DCNV2~\cite{wang2021dcn}: it utilizes kernel product to model vector-wise feature interactions and achieves state-of-the-art performance. 

$\bullet$ FiBiNet~\cite{huang2019fibinet}: it uses SENet and bi-linear operation to improve pair-wise feature interactions. 

$\bullet$ AutoInt~\cite{song2019autoint}: it uses multi-head self-attentive mechanism to learn implicit feature interactions. AutoInt+ improves AutoInt by combining a feed-forward neural network. 

$\bullet$ FiGNN~\cite{li2019fi}: it uses GRU components to model implicit feature interactions in a fully-connected graph. 

$\bullet$ GraphFM~\cite{li2021graphfm}: it performs graph sampling and utilizes multi-head attentive mechanism to model implicit feature interactions in a two-stage manner. 

$\bullet$ ECKD~\cite{zhu2020ensembled}: it is a knowledge distillation based approach, and utilizes large-scale MLP as student network to distill the knowledge from several state-of-the-art teachers.

The above approaches could cover different types in the area of CTR prediction. FmFM, FwFM, xDeepFM and DCNV2 model explicit feature interactions. 
The rest methods model implicit feature interactions. Specially, GraphFM and FiGNN are GNN based methods and ECKD is a knowledge distillation based approach. 
Our proposed KD-DAGFM is a knowledge distillation model based on the field-aware graph neural networks, and is able to capture the high-order explicit feature interactions. 

\subsubsection{Implementation Details} \label{sec: impd}
For each method, grid search is applied to find the optimal settings. All methods are implemented in Pytorch \cite{paszke2019pytorch}.
The dimension of feature embedding is 16. 
The learning rate is in [1e-3, 1e-4, 1e-5]. 
The $L_2$ penalty weight is 1e-5. 
The optimizer is Adam. 
The distillation loss weights $\alpha$ is in [1e-1, 1, 10] and $\beta$ is in [10, 100, 1000].
The depth of CIN and CrossNet is in [2, 3, 4].
The layer size of CIN is in [100, 200, 400].
For DAGFM, the number of propagation layers is in  [2, 3, 4].
For FwFM and FmFM, we adopt field-wise linear weight.
For AutoInt+, the depth, number of head and attention size is 2, 2, 40 respectively.
For ECKD, we choose AutoInt+, DCNV2 and xDeepFM as teachers and adopt "soft label+pre-train" scheme. 
For KD-DAGFM, we report the results with the simple CIN and CrossNet as teachers and the number of  propagation layers of DAGFM is equal to the depth of teacher.

\subsubsection{The Experimental Results of Different CTR Methods.}
We present the results of different models for CTR prediction task in Table~\ref{tab:overall}. We have the following observations:

(1) Deep methods (\ie xDeepFM, DCNV2, AutoInt+) outperform the shallow explicit models (\ie FwFM, FmFM) on the public datasets, which indicates the importance of modeling high-order feature interactions. 

(2) Compared with the implicit methods (\ie AutoInt+, FiGNN, FiBiNet), explicit methods (\ie xDeepFM and DCNV2) achieve better performance on Criteo and MovieLens-1M datasets, which suggests the explicit feature interactions are important patterns for CTR prediction task. As for the datasets Avazu and WeChat, the performance of explicit methods loses the competition with implicit models, due to the over-fitting problem with high sparsity of features.


(3) Most graph-based models (\ie FiGNN, DAGFM) except for GraphFM perform competitively on Criteo, Avazu and Movielens-1M datasets, which suggests that the performance of graph-based model is also influenced by the different aggregation strategies and graph topology. 

(4) For the knowledge distillation model ECKD, by transferring knowledge from multiple teachers, it achieves better performance than graph-based methods, showing the effectiveness of knowledge distillation. 

(5) Our proposed model KD-DAGFM achieves the best performance with knowledge distillation and fine-tuning. KD-DAGFM with knowledge distillation performs better than DAGFM, showing the effectiveness of learning high-order feature interaction from complex teacher models.
The fine-tuning of KD-DAGFM improves the model performance due to the alleviation of the under-fitting problem of the student's feature embedding. 

(6) Different interaction learning functions in KD-DAGFM have different effects: DAGFM-inner outperforms DAGFM-outer on Avazu and WeChat datasets, and loses the advantages on Criteo and Movielens datasets.  

(7) Comparing the number of parameters and FLOPs, we can see that FwFM is the most efficient algorithm due to its simplicity. 
The DNN based methods are much more time-consuming due to their complicated feature interacting units, which makes them impractical in large-scale recommendation scenarios.
In contrast, DAGFM is an efficient model, in which the amount of parameters is comparable with that in the shallow linear methods. It could greatly reduce the computational costs (i.e., FLOPs) by more than 10 times compared with the teacher model, besides, the latency of DAGFM is much more smaller than the other models aside from the light model FwFM, making it possible to be applied into large-scale industry recommender systems.

To summarize, our proposed KD-DAGFM could effectively learn the high-order feature interactions. The fewer parameters make it much more efficient to be applied into large-scale recommendation scenarios.

\subsection{Experimental Analysis of KD-DAGFM}
We analyze the factors that influence the model performance and conduct online A/B testing to show the superiority of KD-DAGFM in real industry recommender systems. 
Besides, an advanced knowledge distillation model KD-DAGFM+ is proposed, 
which improves its ability to be applied
into a wider range of applications.

\subsubsection
{The Effect of Number of Propagation Layers} 
\begin{figure}[!h]
  \centering
  \subcaptionbox{Performance}{
    \includegraphics[width=0.5\linewidth]{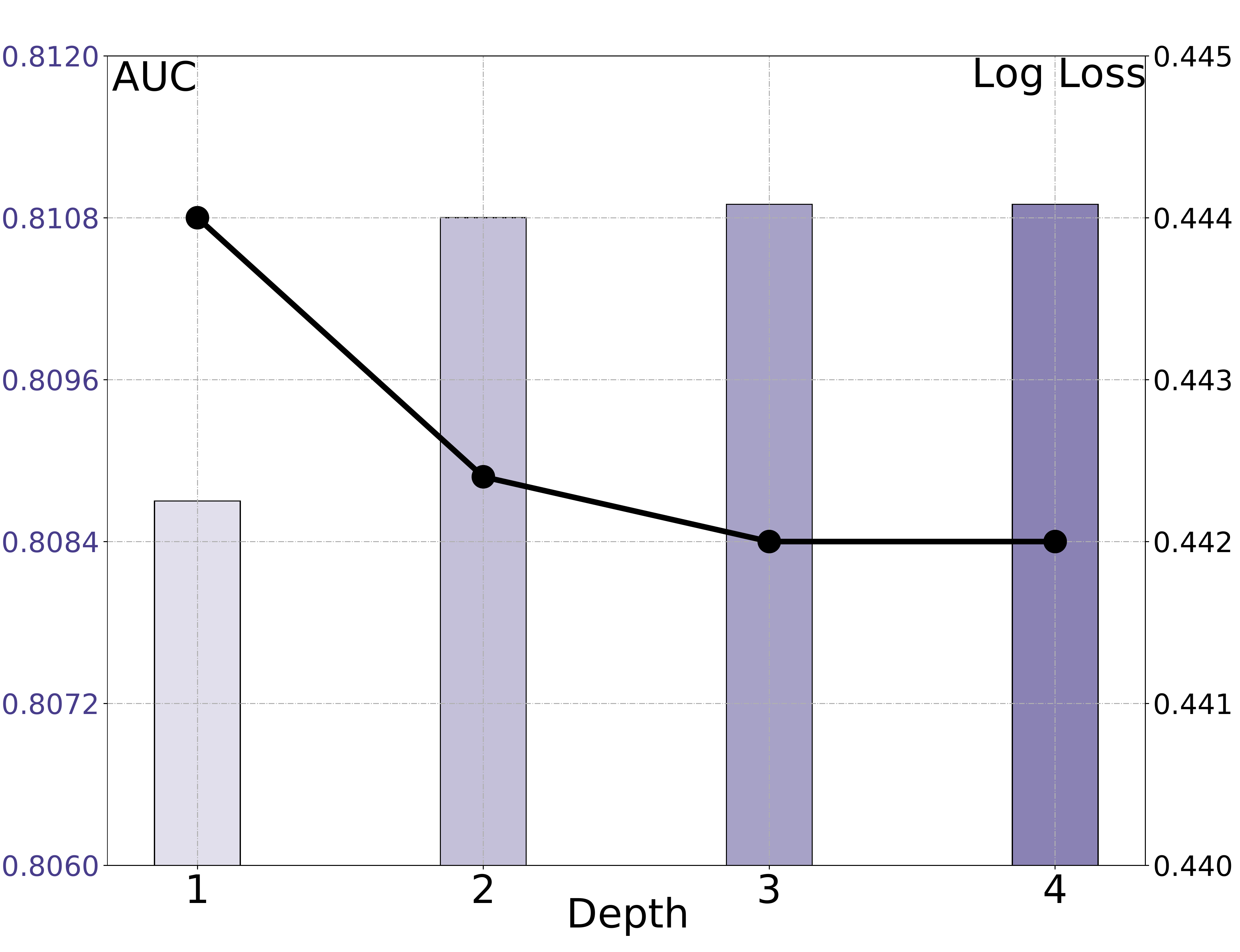}
  }
  \subcaptionbox{KD Loss}{
    \includegraphics[width=0.45\linewidth]{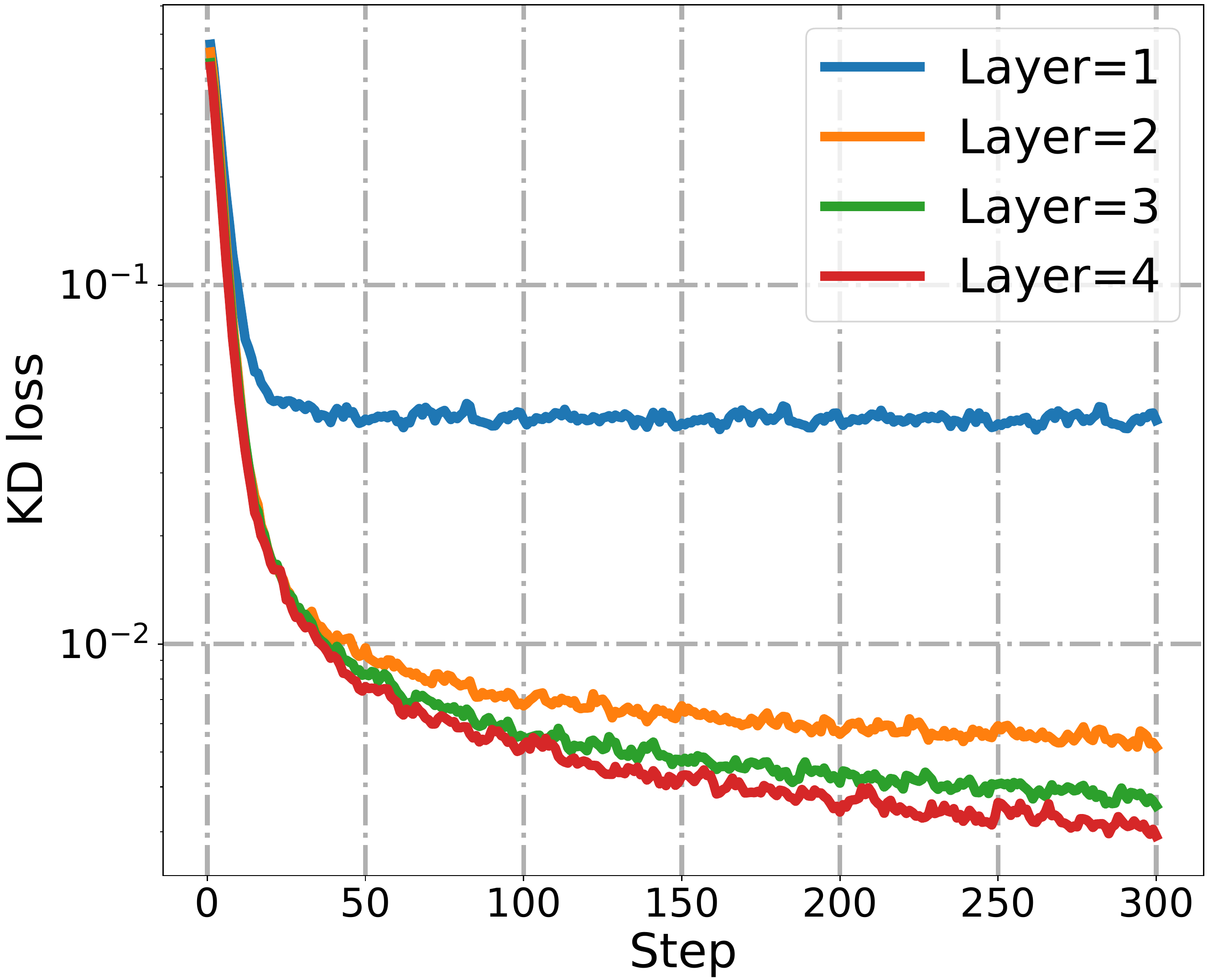}
  }
  \captionsetup{font={small}}
  \caption{The performance of KD-DAGFM with different number of propagation layers.}
  \label{fig:depcom}
\vspace{-1em}
\end{figure}

We explore the effects of the number of propagation layers, which implies the orders of feature interactions.
We use CIN as the teacher network with 3 layers operation (4-order interactions), and the model performance on Criteo dataset is shown in Fig.~\ref{fig:depcom}. 
We can observe that as the number of layers increases, the model performance becomes better, showing the importance of learning the high-order feature interaction information for CTR prediction. 
The improvements of the model with $2$ propagation layers (modeling $3$-order interactions) are significant compared with the one layer model (modeling 2-order interactions), consequently becoming slight as the layers increase, showing the crucial of $2$-order and $3$-order interactions.

\subsubsection{The Effect of Different Interaction Learning Functions}\label{sec: expif} 
\begin{figure}[!h]
  \centering
  \includegraphics[width=1.0\linewidth]{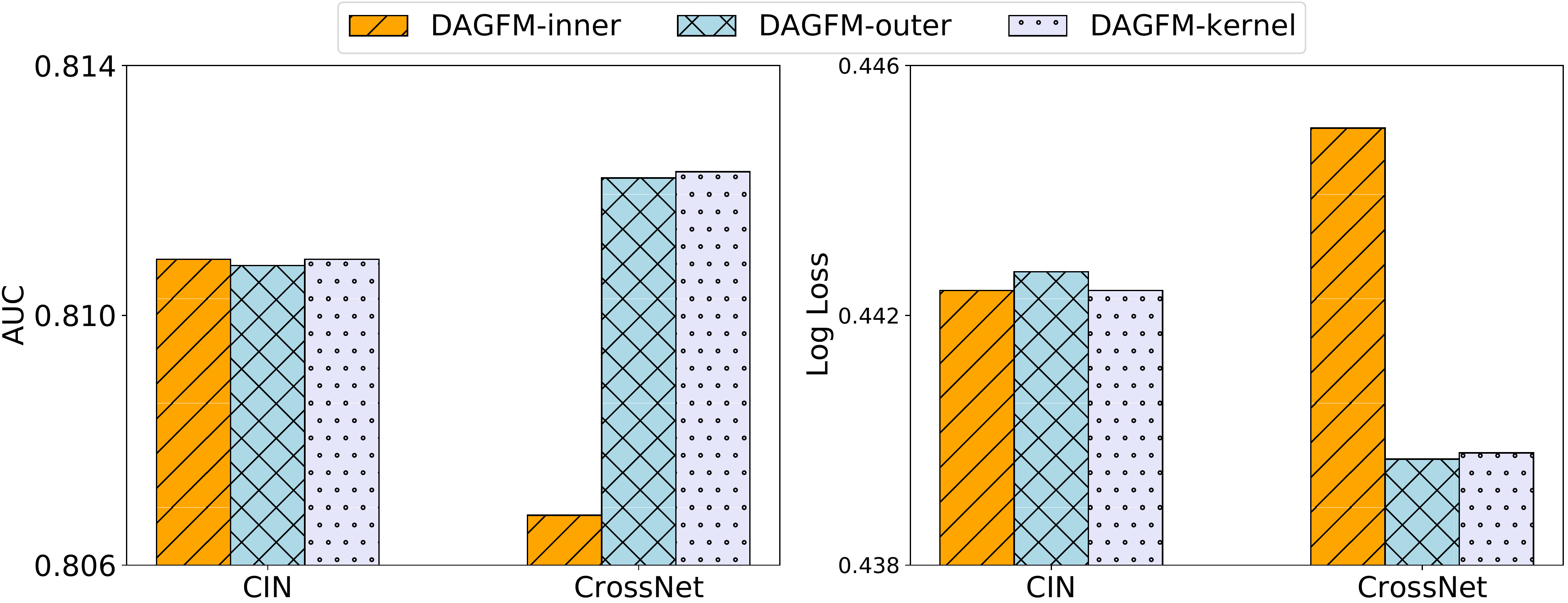}
  \captionsetup{font={small}}
  \caption{The performance of KD-DAGFM with different interaction learning functions.}
  \label{fig:Op}
\end{figure}

As the performance of knowledge distillation model is influenced by the interaction learning functions, we conduct experiments to explore the correlation between teacher models and the interaction learning functions in KD-DAGFM.
As shown in Fig.~\ref{fig:Op}, we can see that DAGFM with our proposed outer learning function can be well suitable for different teacher models (\ie CIN and CrossNet with inner and kernel interaction learning function respectively), while the DAGFM with inner function performs worse with teacher model CrossNet.
Our proposed outer function is decomposed from kernel function, which can be degenerated into inner function, making it more effective to transfer knowledge from the teacher models. 

\subsubsection{Online A/B Testing}
We conduct online experiments on  a large scale industry recommender systems \ie WeChat Official Account Platform\footnote{https://mp.weixin.qq.com/}.
We calculate the gain of the corresponding evaluation metrics: $\#$ Click Users: the number of users had clicked the recommended articles;  CTR: the Click Through Rate.
The SOTA model is an extended version of DCNV2 which achieves the best performance in online service.
Our model KD-DAGFM learns from the CrossNet (as teacher network) with outer interaction learning function.
From Table~\ref{tab:online}, we can see that: 
compared with the SOTA method, KD-DAGFM gains in both the number of click users and the CTR. More importantly, KD-DAGFM outperforms the SOTA method with only 21.5\% computational costs and its inference speed is increased by nearly 3 times, showing the superiority of KD-DAGFM to deal with the web-scale data in real industry recommender systems.

\begin{table}[]
  \captionsetup{font={small}}
  \caption{Online A/B test results on WeChat Official Account Platform. Higher $\uparrow$ is better for Click Users and CTR, while lower $\downarrow$ is better for FLOPs and Latency.} 
  \label{tab:online}
  \resizebox{.4\textwidth}{!}{
  \begin{tabular}{c|cccc}
    \toprule
    Method &\#Click Users $\uparrow$ & CTR $\uparrow$ & FLOPs $\downarrow$ & Latency $\downarrow$\\
    \midrule
    \hline
  SOTA Method & 1,532,276 & 0.09789 & 70M & 0.158 ms\\
    KD-DAGFM & 1,533,351 & 0.09847 & 15M & 0.059 ms\\
    \hline
    Improvements & +0.07\% & +0.59\%   & +78.5\% & +62.7\%\\ 
  \bottomrule
\end{tabular}}
\end{table}

\subsubsection{A General Distillation Model KD-DAGFM+} \label{sec: generalkd}
Considering the teacher models are complex, \ie learning both explicit and implicit feature interactions, in  order to effectively extract the knowledge from both explicit and implicit feature interactions, we propose an advanced distillation model KD-DAGFM+, in which an MLP component are attached after the student model DAGFM (the representations of all node states in DAG is concatenated to feed into the MLP).
Four widely used methods \ie xDeepFM, DCNV2, AutoInt+ and FiBiNet, are chosen as teacher networks, and the results are shown in Table~\ref{tab:gen kd}. We can observe that KD-DAGFM+ achieves approximately lossless distillation performance and effectively improves performance during fine-tuning. 
Besides, comparing with the complex teacher networks, the inference latency of KD-DAGFM+ is reduced greatly.

\begin{table}[!tp]
  \captionsetup{font={small}}
  \caption{Distillation performance of KD-DAGFM+.}
  \label{tab:gen kd}
  \resizebox{.45\textwidth}{!}{
  \begin{tabular}{l|ccc|ccc}
    \toprule
    \multirow{2}{*}{\textbf{ Distillation}}&\multicolumn{3}{c|}{\textbf{Criteo}}&\multicolumn{3}{c}{\textbf{Avazu}}\\
    & AUC & Log Loss & Latency & AUC & Log Loss & Latency \\
    \hline \hline
    xDeepFM (teacher) & 0.8122 & 0.4407& 0.251 ms &  0.7821 & 0.3799& 0.067 ms\\
    KD-DAGFM+ & 0.8122 & 0.4408 & \multirow{2}{*}{0.013 ms} & 0.7821 & 0.3801 & \multirow{2}{*}{0.011 ms} \\
    KD-DAGFM$_{FT}$+ & 0.8132 & 0.4388 & & 0.7870 & 0.3776 \\
    \hline
    DCNV2 (teacher) & 0.8127 & 0.4394 & 0.021 ms & 0.7838 & 0.3782 & 0.013 ms \\
    KD-DAGFM+ & 0.8126 & 0.4396 & \multirow{2}{*}{0.013 ms} & 0.7838 & 0.3784 & \multirow{2}{*}{0.007 ms} \\
    KD-DAGFM$_{FT}$+ & 0.8134 & 0.4387 & & 0.7865 & 0.3775 \\
    \hline
    AutoInt+ (teacher) & 0.8126 & 0.4396 & 0.571 ms & 0.7832 & 0.3786 & 0.332 ms\\
    KD-DAGFM+   & 0.8126 & 0.4396 & \multirow{2}{*}{0.013 ms} & 0.7831 & 0.3784 & \multirow{2}{*}{0.010 ms} \\
    KD-DAGFM$_{FT}$+ & 0.8137 & 0.4385 & &  0.7875 & 0.3761 \\
    \hline
    FiBiNet (teacher) & 0.8126 & 0.4415 & 0.124 ms & 0.7837 & 0.3783 & 0.024 ms\\
    KD-DAGFM+ & 0.8125 & 0.4418 & \multirow{2}{*}{0.009 ms} & 0.7836 & 0.3786 & \multirow{2}{*}{0.008 ms} \\
    KD-DAGFM$_{FT}$+ & 0.8131 & 0.4405 & &  0.7875 & 0.3768 \\
  \bottomrule
\end{tabular}}
\end{table}
\section{Related work}

\paratitle{Feature Interactions Learning.}
Learning feature interactions is a fundamental problem in CTR prediction tasks. 
FM~\cite{rendle2010factorization} is the most basic and widely used model, which assigns a learnable embedding vector to 
each feature to capture the second-order feature interactions.
Besides, HOFM~\cite{blondel2016higher} captures the high-order feature interactions; 
FwFM~\cite{pan2018field} and FmFM~\cite{sun2021fm2} utilize inner product and kernel product to capture field interaction respectively. 
To better model high order feature interactions, lots of deep learning based methods learn feature interactions through a well-designed component and incorporate a deep neural network.
For example, DeepFM~\cite{guo2017deepfm} imposes a factorization machine as explicit component; xDeepFM~\cite{lian2018xdeepfm} and DCNV2~\cite{wang2021dcn} propose the Compressed Interaction Network (CIN) and Cross Network to model the explicit feature interactions.
Besides, 
recent research~\cite{zhu2021aim,khawar2020autofeature,liu2020autofis,liu2020autogroup,chen2019bayesian,zhao2020amer,song2020towards}
trend focuses on learning effective feature interactions via AutoML.
However, these deep models have extremely high computational complexity, making them impractical be applied to real-time large-scale recommendation scenarios.

\paratitle{Graph Neural Networks.}
In recent years, Graph Neural Network is widely used in recommender systems. 
Many GNN-based studies~\cite{wang2019neural,he2020lightgcn,ying2018graph,wu2021self} capture the collaborative signal existing in the user-item bipartite graph and treats the recommendation task as a link prediction problem. 
In the area of GNN based CTR predictions, DG-ENN \cite{guo2021dual} exploits the strengths of graph representation with two  carefully designed learning strategies to refine the feature embeddings; Fi-GNN \cite{li2019fi} learns implicit feature interactions by graph propagation on a fully-connected graph; GraphFM \cite{li2021graphfm} proposes a two-stage aggregation strategy to model high-order feature interactions;
PCF-GNN \cite{li2021explicit} proposes a pre-trained model to generate the explicit semantic cross features.
Interestingly, a recent study \cite{dudzik2022graph} points out that GNNs are claimed to align with dynamic programming.
In this work, we propose a lightweight and directed acyclic graph based model to learn the explicit high-order feature interactions. 

\paratitle{Knowledge Distillation.}
Knowledge Distillation (KD)~\cite{hinton2015distilling} is a method to extract knowledge from sophisticated models and compress it into a simple model, so that it can be deployed to practical applications. 
As KD has achieved great success in many research fields such as natural language processing~\cite{tan2019multilingual},
the KD based models~\cite{zhu2020ensembled,zhou2018rocket,xu2020privileged} have attracted increasing attention in recommender systems.
ECKD~\cite{zhu2020ensembled} is an ensemble learning approach via knowledge distillation to deliver the performance of MLP. It achieves better performance than its teacher at cost of an undiminished amount of parameters of its student.
For model compression, the general defect of knowledge distillation is the degradation of model accuracy because the capacity gap between large teachers and small students always exists. The design of an effective student model is still a challenge.
\section{Conclusion}
In this paper, we propose a lightweight knowledge distillation model KD-DAGFM with a directed acyclic graph neural network for CTR predictions. 
The directed acyclic graph structure proposed in KD-DAGFM makes it possible to align with a DP algorithm, which improves its capability to learn arbitrary feature interactions from the complex teacher networks and makes it  achieve approximately lossless model performance.
The information propagation in the student model DAGFM with different interaction learning function could greatly reduce the computational resource costs. KD-DAGFM achieves the best performance on both online and offline experiments, showing the superiority of DAGFM to deal with the industrial scale data in CTR prediction task.




\begin{acks}
{This work is supported by the National Natural Science Foundation of China under Grant No. 62102038 and 62222215, and Weixin Open Platform under Grant No. S2021120.}
\end{acks}

\bibliographystyle{ACM-Reference-Format}
\balance
\bibliography{sample-base}


\end{document}